\documentclass[12pt]{article}
\usepackage{epsfig}
\begin{document}
\newcommand{\bq}{\begin{equation}}
\newcommand{\eq}{\end{equation}}
\newcommand{\bqa}{\begin{eqnarray}}
\newcommand{\eqa}{\end{eqnarray}}
\newcommand{\nl}{\nonumber \\}
\newcommand{\f}{\varphi}
\newcommand{\suml}{\sum\limits}
\newcommand{\plaat}[4]{\raisebox{#4pt}{\epsfig{figure=#1.eps,
  width=#2cm,height=#3cm}}}
\newcommand{\cint}[1]{\int\limits_{#1}}
\newcommand{\Hs}{\mbox{``}H\mbox{''}(\phi^s)}
\newcommand{\Hsk}{\mbox{``}H\mbox{''}(\phi_k^s)}

\begin{center}
{\bf {\Large Vertex counting:\\
Statistical Distribution of Vertices\\ in Large Sets of Tree Diagrams}\\
\vspace*{\baselineskip}
Petros Draggiotis\footnote{{\tt petros@sci.kun.nl}}
 and Ronald Kleiss\footnote{{\tt kleiss@sci.kun.nl}}\\
University of Nijmegen, the Netherlands}\\
\vspace*{3\baselineskip}
Abstract
\end{center}
We study the problem of determining the distribution of vertices of 
a particular given type in the set of all Feynman tree graphs in
quantum field theories. We show that in almost all cases
a Gaussian distribution arises asymptotically, and we compute the mean and
variance of this distribution for several theories. We show
the distribution's `fine structure', arising from
topological sum rules, can be obtained.

\newpage
\section{Introduction}
Recently, computational prowess in particle phenomenology has progressed
to the evaluation of multi-particle scattering amplitudes, at least at
the tree level, with ten or more external legs. The concomitant enormous
number of Feynman graphs ({\it e.g.} 10.5 million for gluonic $2\to8$
scattering) suggests the study of the Feynman graphs themselves as an 
ensemble of combinatorial objects. The number of graphs for various processes
in self-interacting theories
has been studied in a number of publications \cite{AKPetc,Vol}, and also results
for theories involving different particles have been found \cite{DKetc}.
It is therefore natural to examine various statistical aspects of these
ensembles more closely. In the present paper we deal with the problem of
determining the frequecny of occurrence of particular vertices. For instance,
one may wish to find the frequency of occurrence of four-gluon vertices
in the ensemble of Feynman graphs for gluonic QCD: it is this kind
of question that we address in this paper.
The layout of the paper is as follows. First, we
show that the frequency distribution of any kind of vertex tends to 
a Gaussian distribution as the number of external lines becomes asymptotically 
large. We then proceed to evaluate the parameters of these distributions
for several theories, notably gluonic QCD and theories with all possible
vertices present. These results hold for the `coarse-grained' point of view 
in which the number of vertices is interpreted as a continuous,
rather than a discrete, fraction of the number of legs. Subsequently, we
discuss how the `fine structure' of the distribution can be determined:
this fine structure arises from the topological conservation laws that
govern the number of vertices in any tree amplitude with a given number
of legs. Finally, we turn to the more realistic case of QCD, in which 
both quark and gluon fields occur.

\section{Gaussian limits for vertex occurrence}
Consider a self-interacting theory of a single field $\f$. Since we will only be
counting diagrams and vertices, the particular types of interactions are
only important in terms of the number of legs involved in each vertex.
We shall start with a quick review of the counting of diagrams.
Let the theory have vertices of type $\f^{q+1}$ for some set of numbers 
$q\ge2$ (that is, we allow any combination of 3,4,5,$\ldots$ vertices).
We define the `potential' 
\bq
W(\f) = \suml_{q\ge2}{\epsilon_q\over q!}\f^q
\eq
where $\epsilon_q=1$ if the vertex occurs in the action (or effective
action), otherwise it is zero. If we denote the tree-level $1\to n$
amplitude \footnote{Throughout this paper, the `amplitude' is defined as the number
of diagrams} by $a(n)$, we can define the amplitude-generating function
by
\bq
\phi(x) = \suml_{n\ge1} {x^n\over n!}a(n)
\eq
Then, simple diagrammatic arguments \cite{AKPetc} show that the 
Schwinger-Dyson equation for the number of diagrams is
\bq
\phi(x) = x + W(\phi(x))
\label{SDeqn}\eq
Iteration of this relation, where $\phi(x)$ is a power series in $x$,
then gives the number of graphs for any desired value of $n$; alternatively,
the analytical structure of $\phi(x)$ as a function of $x$ informs us about
the asymptotic behaviour of $a(n)$ with $n$. 
In particular, it is clear that in the general case $\phi(x)$ will be singular
in $x$, and moreover that this singularity will be a branch cut rather
than a pole. The condition for the singularity is, of course
\bq
\left|{d\phi(x)\over d x}\right|=\infty\;\;\Rightarrow\;\;
{d x\over d\phi}=0\;\;\Rightarrow\;\;W'(\phi)=1
\eq
Since $q\ge2$ there are in general several solutions $\phi_0$ to this 
condition, and the radius of convergence of $\phi(x)$ as a power series
is given by that solution $\phi_0$ which leads to the smallest absolute
value of $x=\phi_0-W(\phi_0)$. Taking this value of $\phi_0$, we see that,
asymptotically, the number of graphs goes as
\bq
a(n) \sim {n!\over(\phi_0-W(\phi_0))^n}\;\;,
\label{crude}\eq
multiplied by subleading terms: in fact, the more precise form has been
derived in \cite{AKPetc} to read
\bq
a(n) \sim {n!\over n^{3/2}}{1\over(\phi_0-W(\phi_0))^n}
\sqrt{{\phi_0-W(\phi_0)\over2\pi W''(\phi_0)}}
\label{precise}\eq
but for the moment the leading form (\ref{crude}) is sufficient.

We now turn to the counting of vertices. Let us focus on the vertex of type
$\f^{q+1}$ for some $q$ with $\epsilon_q=1$. We can count the occurrence
of this vertex by giving it a weight $z$ in the Schwinger-Dyson equation
(\ref{SDeqn}), which then becomes
\bq
\phi(x,z) = x + W(\phi(x,z)) + (z-1){\phi(x,z)^q\over q!}
\label{SDmod}\eq
The amplitude-generating function now depends on $z$ as well as on $x$. 
By iteration of this implicit equation we can find
\bq
\phi(x,z) = \suml_{n\ge1}{x^n\over n!}a(n;z)
\eq
to desired order in $x$. In the coefficient
\bq
a(n;z) = \suml_{m\ge0} b(n;m)z^m
\eq
which is a finite polynomial in $z$, the number $b(n;m)$ is then the
number of $1\to n$ graphs with precisely $m$ vertices of type $\f^{q+1}$;
and $a(n;1)$ is the total number of graphs.
It is more instructive to consider not the number of graphs themselves
but rather
the probability to pick a diagram with $m$ $(q+1)$-vertices at random
out of all tree graphs in the $1\to n$ amplitude.
If we denote this probability by $\pi(n;m)$, we have simply
\bq
P_n(z) \equiv \suml_{m\ge0} \pi(n;m)z^m = {a(n;z)\over a(n;1)}\;\;.
\eq
Putting in the leading asymptotic behaviour for $a(n)$ of Eq.(\ref{crude})
we therefore find that
\bq
\suml_{m\ge0} \pi(n;m)z^m \sim \left({F(\phi_0(1);1)\over 
F(\phi_0(z);z)}\right)^n\;\;\;,\;\;\;
F(\f;z) \equiv \f - W(\f) - (z-1){\f^q\over q!}
\eq
and $\phi_0(z)$ is of course the value of $\phi(x;z)$ at the singularity
nearest to the origin in the complex $x$ plane:
\bq
W'(\phi_0(z)) - (z-1){\phi_0(z)^{q-1}\over(q-1)!} = 1\;\;.
\eq
The distribution of the $m$ values is determined once we know its
moment-generating (or characteristic) function. It is simply related
to the result we have obtained: 
\bqa
\left\langle e^{m\xi}\right\rangle &\equiv&
\suml_{k\ge0}{\xi^k\over k!}\left\langle m^k\right\rangle
= \suml_{m,k\ge0}\;\pi(n;m)\;{m^k\xi^k\over k!}\nl
&=& \suml_{m\ge0}\;\pi(n;m)\;(e^\xi)^m
= \left({F(\phi_0(1);1)\over F(\phi_0(e^\xi);e^\xi)}\right)^n
\eqa
But, by elementary arguments from probability theory, 
this implies that $m$ is distributed as the sum of $n$ independent,
identically distributed random variables. The central limit theorem therefore
asserts (except in special cases to which we shall come later) that
{\bf for large $n$, the values of $m$ are normally distributed}:
\bq
\pi(n;m) \sim {1\over\sqrt{2\pi n\sigma^2}}
\exp\left(-{n\over2\sigma(q)^2}\left({m\over n}-\mu(q)\right)^2\right)
\eq
with the parameters $\mu(q)$ and $\sigma(q)^2$ still to be determined
(note that since $m$ is always integer, it is rather the {\em ratio}
$m/n$ that follows the continuous normal distribution).

\section{Results for various theories}
Having established that the $m$ values are normally distributed, we need only
to compute the mean and variance.
We have, for the first two factorial moments,
\bq
\langle m\rangle = P_n'(1)\;\;\;,\;\;\;\langle m(m-1)\rangle = P_n''(1)\;\;.
\eq
Some straightforward algebra then leads us to
\bqa
\mu(q+1) &=& {f^q\over q!(f-W(f))}\;\;,\nl
\sigma(q+1)^2 &=& \mu(1+\mu)-\left({f^{q-1}\over(q-1)!}\right)^2
{1\over W''(f)(f-W(f))}\;\;,
\eqa
where $f=\phi_0(1)$, hence $W'(f)=1$. 
Note that, since all the Taylor coefficients of $W(\f)$
are nonnegative, the numbers $f$, $f-W(f)$ and $W''(f)$ are all positive.
As a further check, we recall the topological relation, valid for
any given tree diagram:
\bq
\suml_{k\ge3} (k-2)m_{k} = n-1
\eq
where $m_{k}$ is the number of vertices of type $\f^{k}$ in that diagram.
Indeed, it is trivially seen that
\bq
\suml_{q\ge2}\;\epsilon_q(q-1)\mu(q+1) = 1\;\;.
\eq
In addition, we can straightforwardly extend our discussion to the case where
we consider the combined distribution of two different
vertex types, with $q_1+1$ and $q_2+1$ legs ($q_1\ne q_2$), 
by introducing two counting weights $z_{1,2}$. The result
for the expected value of the product $m_{q_1+1}m_{q_2+1}$ is, then,
\bq
\langle m_{q_1+1}m_{q_2+1}\rangle = 
{n^2+n\over(f-W(f))^2}{f^{q_1+q_2}\over q_1!q_2!}
-{n\over W''(f)(f_W(f))}{f^{q_1+q_2-2}\over(q_1-1)!(q_2-1!)}\;\;,
\eq 
which provides the additional check
\bqa
&&0 = \suml_{q}\epsilon_q(q-1)^2\sigma(q+1)^2\nl
&&+\suml_{q_1\ne q_2}\epsilon_{q_1}\epsilon_{q_2}(q_1-1)(q_2-1)
\left({\langle m_{q_1+1}m_{q_2+1}\rangle\over n}
-\mu(q_1+1)\mu(q_2+1)\right)\;\;.
\eqa

We now turn to a few concrete cases. In the first place, there is gluonic
QCD, with three- and four-point vertices, so that
\bq
W(\f) = {1\over2}\f^2 + {1\over6}\f^3\;\;.
\eq
For $z=1$, the nearest singularity is at $x=\sqrt{3}-4/3$, reached for
$f=\phi_0(1) = -1+\sqrt{3}$. We find that
\bqa
\mu(3) = {6\sqrt{3}-3\over11} &\;\;,\;\;& 
\sigma(3)^2 = {52\sqrt{3}-48\over121}\;\;\;,\nl
\mu(4) = {7-3\sqrt{3}\over11} &\;\;,\;\;&
\sigma(4)^2 = {13\sqrt{3}-12\over121}\;\;.
\eqa
In figures 1 and 2 we give the frequency distribution of the four-point
vertices for this theory, for $n=25$ (2.3$\;10^{32}$ diagrams)
and $n=55$ (5.5$\;10^{91}$ diagrams).
\begin{figure}
\begin{center}
\plaat{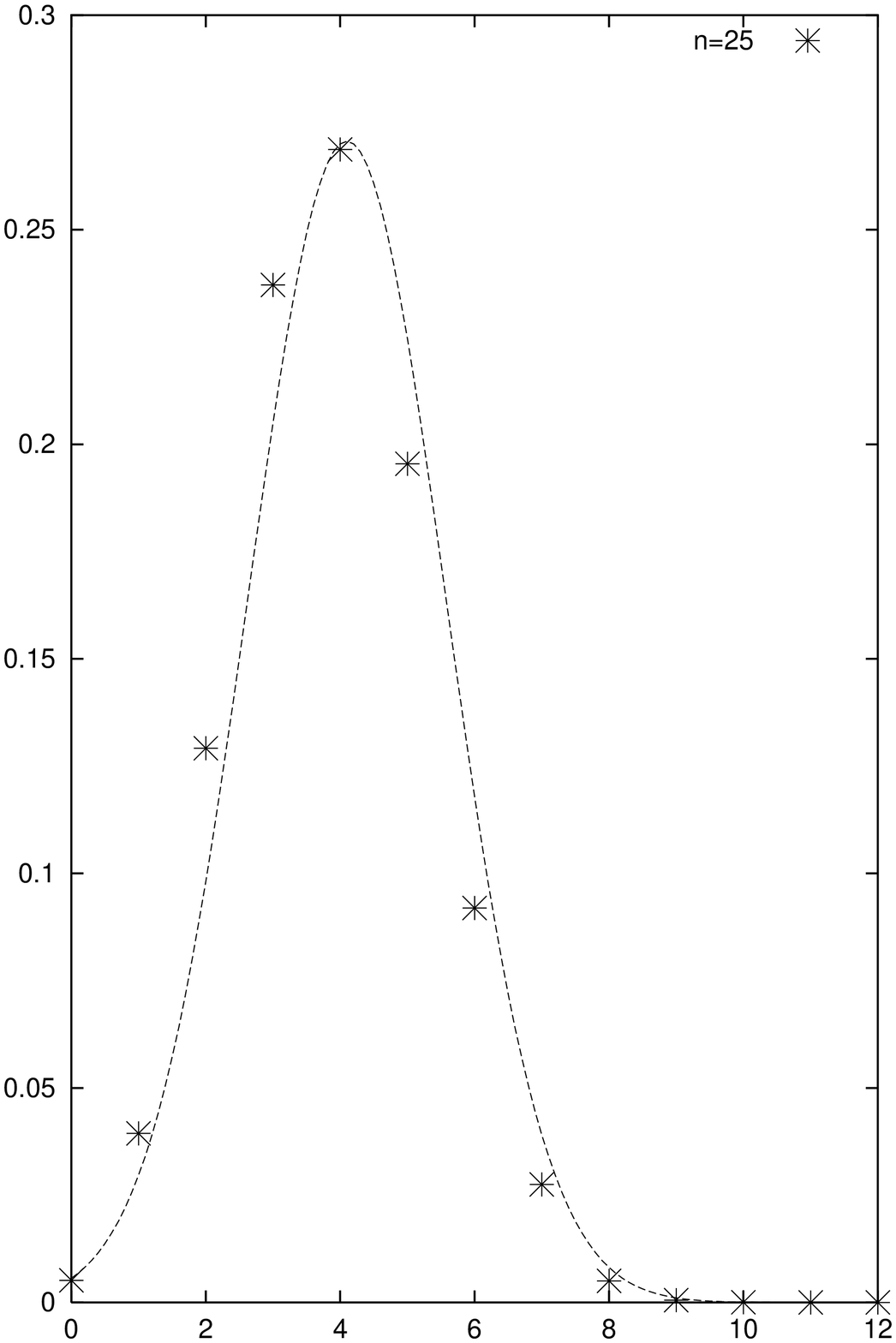}{12}{14}{0}
\caption[.]{Frequency distribution of the number of 4-point vertices
in $\f^3+\f^4$ theory, for $n=25$. The solid line is the asymptotic
estimate.}
\end{center}
\end{figure}
\begin{figure}
\begin{center}
\plaat{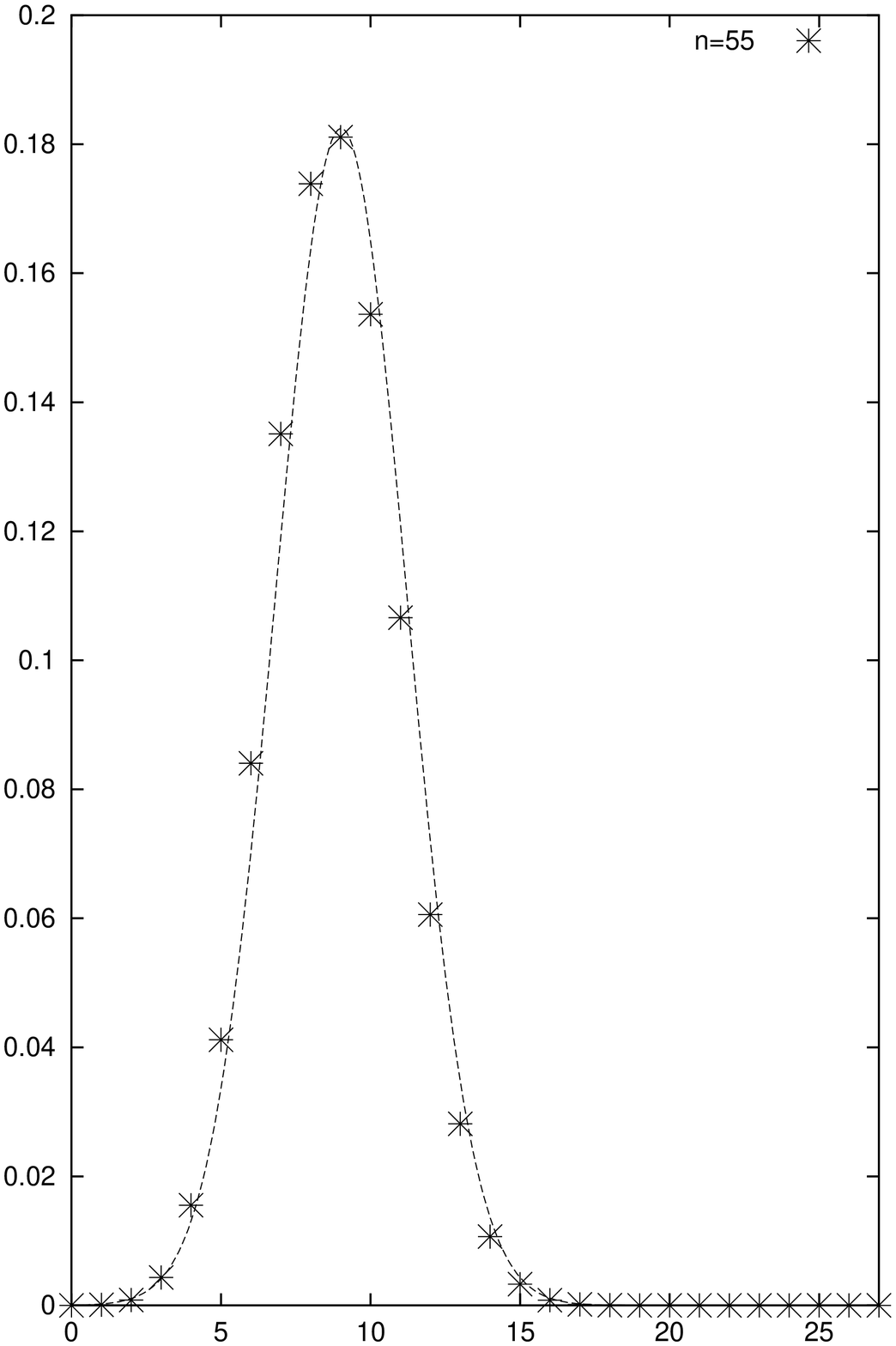}{12}{14}{0}
\caption[.]{Frequency distribution of the number of 4-point vertices
in $\f^3+\f^4$ theory, for $n=55$. The solid line is the asymptotic
estimate.}
\end{center}
\end{figure}
\begin{figure}
\begin{center}
\plaat{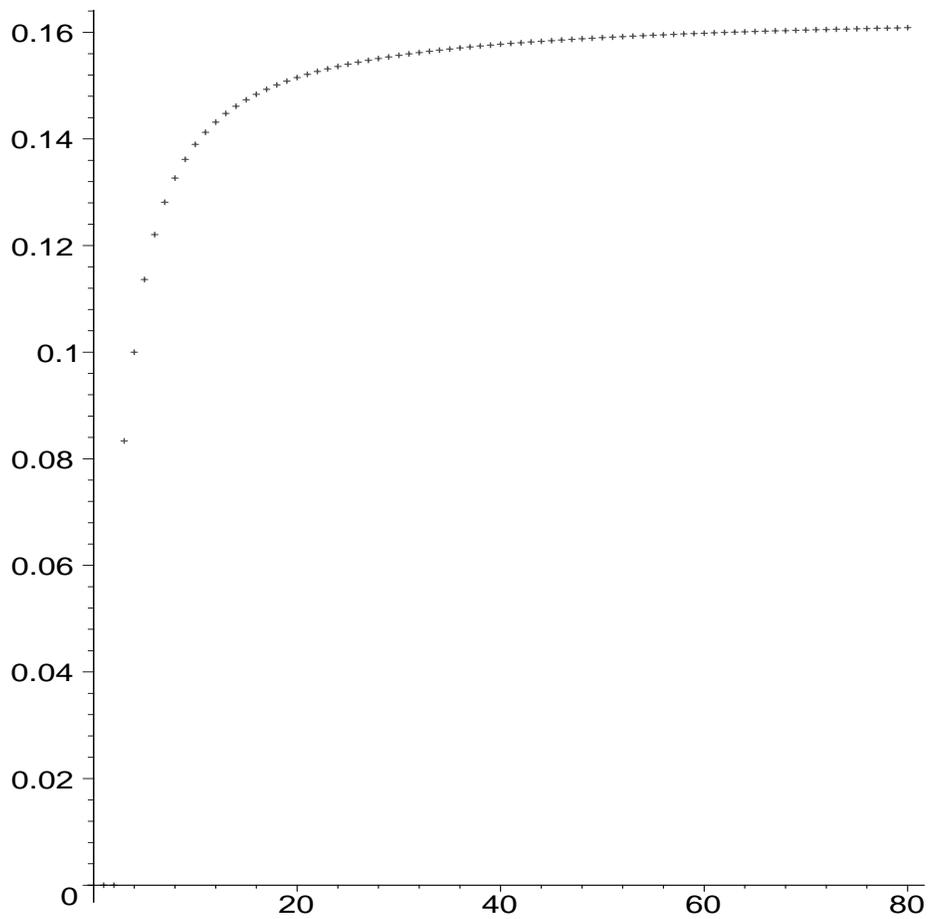}{12}{12}{0}
\caption[.]{Average fraction $\langle m\rangle/n$ for four-point vertices,
as a function of $n$, in $\f^3+\f^4$ theory.}
\end{center}
\end{figure}
In figure 3, we give the actual value of the average number of four-point
vertices as a fraction of $n$, for $1\le n\le80$. The asymptotic value,
0.163986$\ldots$, is reached from below (as already evident from
figures 1 and 2): the correction term appears to go as $1/4n$ in this
range.\\

The second case of interest is a theory (like an effective theory, after tadpole and 
mass renormalization) where
all types of vertices occur, that is, $\epsilon_q=1$ for all $q$.
We have
\bq
W(\f) = e^\f - \f - 1\;\;,
\eq 
so that $f=\log(2)$, $f-W(f)=2\log(2)-1$, and $W''(f)=2$, and we have
\bqa
\mu(q+1) &=& {(\log2)^q\over q!(2\log2-1)}\;\;,\nl
\sigma(q+1)^2 &=& \mu(q+1)\left[
1 + \mu - {(\log2)^{q-2}\over2(q-1)!}\right]\;\;.
\eqa
In figure 4 we plot $\log\mu(q+1)$ for the first few values of $q$ for this
theory. The values of $\sigma(q+1)^2$ follow quite closely: the numerical values
of $\sigma(6)^2$ and $\mu(6)$ already differ by less than one percent.
\begin{figure}
\begin{center}
\plaat{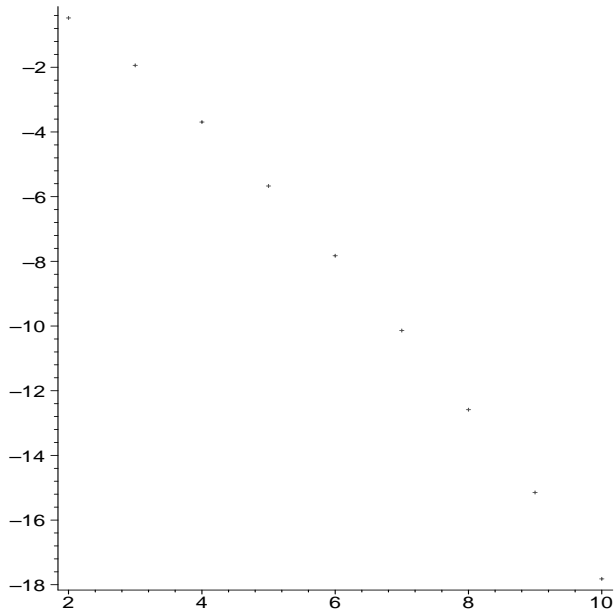}{8}{8}{0}
\caption[.]{$\log\mu(q)$ as a function of $q$ for a theory with
all possible vertices.}
\end{center}
\end{figure}

\section{Fine structure}
In figure 5, we give the actual and asymptotic distribution of 3-point vertices
in $\f^3+\f^4$ theory, for $n=35$. 
\begin{figure}
\begin{center}
\plaat{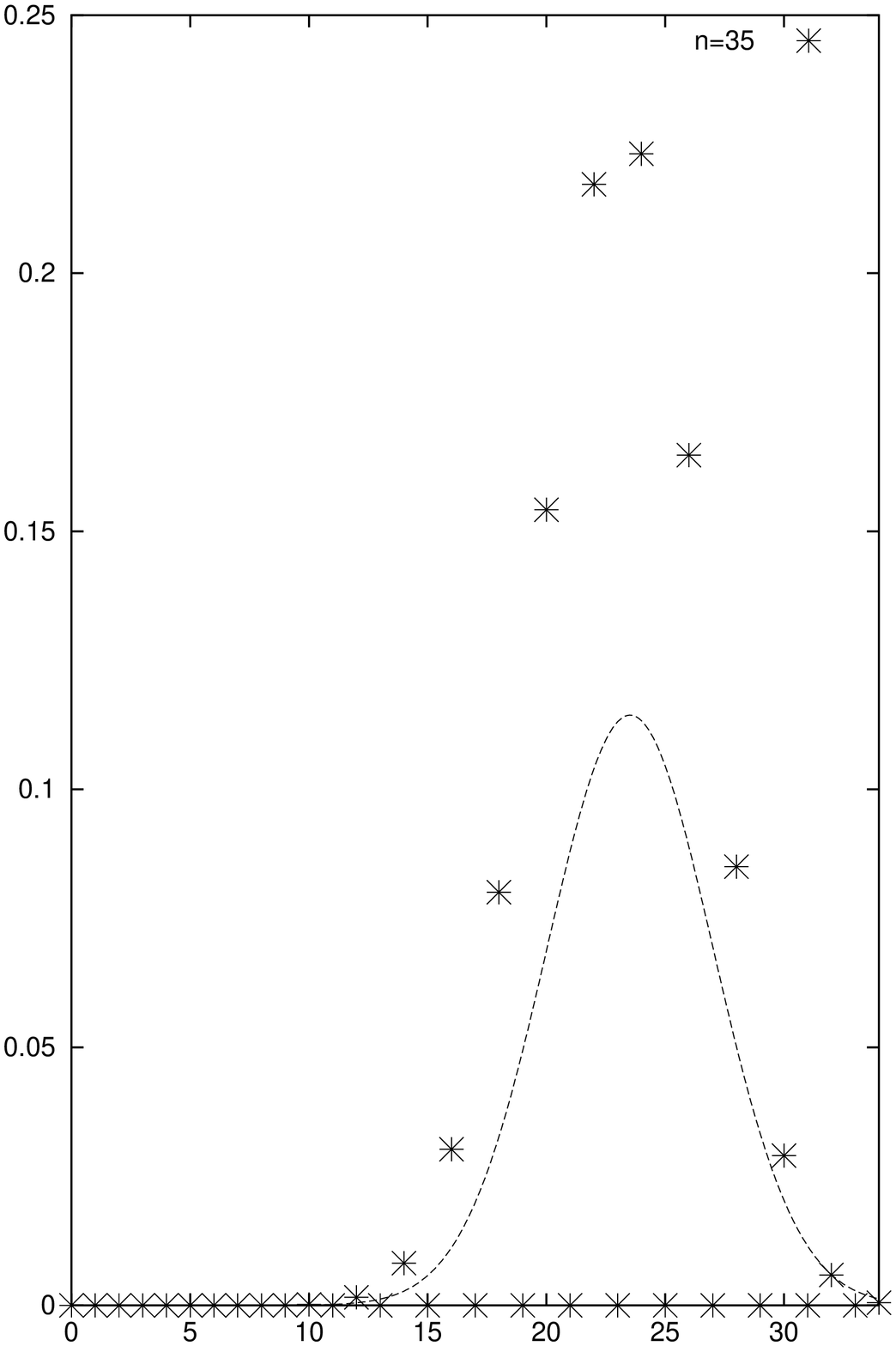}{12}{12}{0}
\caption[.]{Frequency distribution of 3-point vertices in $\f^3+\f^4$
theory, for $n=35$. The solid line is the coarse-grained
asymptotic estimate.}
\end{center}
\end{figure}
Every Feynman diagram contains an even 
number of 3-point vertices. This follows of course from the topological
sum rule for this theory:
\bq
m_3 + 2m_4 = n-1
\eq
so that $m_3$ is either always even or always odd, depending on the parity 
of $n$. The asymptotic estimate does not reflect this; we shall now
describe how also this `fine structure' can be obtained, even in
the asymptotic limit, as a consequence
of the algebraic structure of the action. In order to be slightly more
general, let us assume that there are two types of vertices present, one
with $q+1$ legs and one with $p+1$ legs, and that we count the
number of vertices of the first type. The Schwinger-Dyson equation now reads
\bq
\phi = x + z{\phi^q\over q!} + {\phi^p\over p!}
\eq
and the condition for the singularity reads
\bq
z{\phi^{q-1}\over(q-1)!} + {\phi^{p-1}\over(p-1)!} = 1\;\;.
\eq
After having obtained $a(n;z)$, we can of course extract $b(n;m)$ by standard
means:
\bq
b(n;m) = {1\over2i\pi}\cint{z\sim0}{a(n;z)\over z^{m+1}}dz\;\;,
\eq
where the integral is over an infinitesimal counterclockwise loop around
$z=0$. Now notice that, for $z=0$, the singularity condition has not one
but $p-1$ distinct solutions, which we shall denote by $\phi_k$:
\bq
\phi_k = (s!)^{1/s}\exp\left({2i\pi k\over s}\right)\;\;\;,\;\;\;
k=0,1,2,\ldots,s-1\;\;\;,\;\;\;s\equiv p-1\;\;.
\eq
The corresponding values for $x=\phi_k-\phi_k^p/p!$ all have the same
absolute value (at $z=0$, to be sure) and contribute equally to the
asymptotic result. It therefore makes more sense to perform the
integral not in terms of $z$ but in terms of the $\phi_k$, using
Eq.(\ref{precise}). To this end, we write, for nonzero $z$, the
singularity condition as
\bq
z = {(q-1)!\over\phi^{q-1}}\left(1-{\phi^s\over s!}\right)
= {\Hs\over\phi^{q-1}}
\eq
where by $\Hs$ we denote an unspecified function of $\phi^s$. Inserting this
expression for $z$ everywhere, we find
\bq
F(\phi,z) = \phi\Hs\;\;\;,\;\;\;
{\partial^2\over\partial\phi^2}F(\phi,z) = {\Hs\over\phi}
\;\;\;,\;\;\;{dz\over d\phi} = {\Hs\over\phi^q}\;\;.
\eq
The above form for $b(n;m)$ is now a sum of loop integrals:
\bqa
b(n;m) &\sim& \sum\limits_{k=0}^{s-1}\;{n!\over 2i\pi n^{3/2}}\;
\cint{\phi\sim\phi_k}\;{1\over z^{m+1}F(\phi,z)^n}
\sqrt{{-F(\phi,z)\over 2\pi{\partial^2\over\partial\phi^2}F(\phi,z)}}
{dz\over d\phi}\;d\phi\nl
&=& \sum\limits_{k=0}^{s-1} \;\cint{\phi\sim\phi_k}
 {1\over\phi^{n-m(q-1)}}\Hs\;d\phi\nl
&=& \cint{\phi\sim\phi_0} {1\over\phi^{n-m(q-1)}}\Hs\;d\phi\nl
&&\hphantom{AAAAAAAA}\times
\left(\sum\limits_{k=0}^{s-1}\exp\left(
{2i\pi k\over s}\right)^{1-n+m(q-1)}\right)\;\;,
\eqa
owing to the $s$-fold symmetry of the unspecified function $\Hs$.
Now, the sum of the powers of the roots of unity in the above gives zero
except when $1-n+m(q-1)$ happens to be a multiple of $s=p-1$, or in
other words
\bq
m(q-1) + m'(p-1) = n-1
\eq
for some $m'$, which is again precisely the topological sum rule.
In that case, we find exactly $s$ times the result from a single
singular point, so that the normalization of the asymptotic distribution
is preserved.

Two remarks are in order here. In the first place, the conclusion remains
unchanged if other vertices are present, as long as these all have
$ks+1$ legs ($k$ any integer larger than 1). If this algebraic symmetry
is destroyed by the presence of another vertex type, there will in general be
only one dominant singular point, and the vertex frequency distribution
will not show any `quantisation' anymore. 
In the second place, it is instructive to note that we arrive at the
quantisation condition by working in the neighbourhood of $z=0$: once we
move out to $z=1$, where the singular points are no longer symmetrically
distributed, we move toward coarse-graining and a continuous distribution.

\section{Vertex counting in QCD}
We now turn to the more realistic case of QCD with both quarks and
gluons. For simplicity we shall only consider a single quark flavour, as 
extensions to the case of more flavours is straightforward.
We shall count the various vertices by assigning to the $q\bar{q}g$ vertex
a weight $z_q$, and to the three- and four-gluon vertices weights
$z_3$ and $z_4$, respectively. In any $1\to n$ amplitude we
shall assign a factor $\bar{u}$ to an outgoing quark, a factor $v$ to
an outgoing antiquark, and a factor $x$ to an outgoing gluon as before.
The amplitude-generating function for an incoming quark is denoted by
$\bar{\psi}$, that for an incoming antiquark by $\psi$, and that for
an incoming gluon by $\phi$, again as before. The QCD Feynman rules
now lead to the following coupled Schwinger-Dyson equations at the
tree level:
\bqa
\bar{\psi} &=& \bar{u} + z_q\bar{\psi}\phi\;\;\;,\;\;\;
\psi = v + z_q\psi\phi\;\;,\nl
\phi &=& x + {z_3\over2}\phi^2 + {z_4\over6}\phi^3
+ z_q\bar{\psi}\psi\;\;,
\eqa
We can rewrite this as an equation in terms of $\phi$ alone:
\bq
\phi = x + {z_3\over2}\phi^2 + {z_4\over6}\phi^3
+ {z_q\xi\over(1-z_q\phi)^2}\;\;\;,\;\;\;\xi\equiv\bar{u}v\;\;.
\eq
The occurrence of the combination $\xi$ reflects fermion number
conservation. The generating function $\phi$ is
\bq
\phi = \phi(x,\xi,\vec{z}) = 
\suml_{n_g\ge0}\suml_{n_q\ge0}\suml_{m_3\ge0}\suml_{m_4\ge0}
\suml_{m_q\ge0}c(n_g,n_q;m_3,m_4,m_q)
{x^{n_g}\over n_g!}{\xi^{n_q}\over n_q!^2}z_3^{m_3}z_4^{m_4}z_q^{m_q}
\eq
where $\vec{z}=(z_3,z_4,z_q)$, and
$c(n_g,n_q;m_3,m_4,m_q)$ denotes the number of Feynman tree
graphs with one incoming gluon, $n_g$ outgoing gluons, $n_q$ outgoing
$q\bar{q}$ pairs, and precisely $m_3$ three-gluon vertices, 
$m_4$ four-gluon vertices and $m_q$ quark-gluon vertices.
Again, explicit iteration of this equation gives $\phi$ as a multinomial in
all its arguments, from which the number of graphs can be read off easily.
Due to the larger number of counting variables the arriving at asymptotic
estimates is more cumbersome in this case, but not qualitatively different.
We have
\bqa
x &=& F(\phi,\xi,\vec{z})\nl &=&  
\phi - {z_3\over2}\phi^2 - {z_4\over6}\phi^3
- {z_q\xi\over(1-z_q\phi)^2}\;\;,
\eqa
and the large-$n_g$ behaviour is determined as before by requiring $dx/d\phi$
to vanish, which occurs when
\bq
\xi = \Xi(\phi,\vec{z}) = 
{1\over2z_q^2}(1-z_q\phi)^3\left(1-z_3\phi-{z_4\over2}\phi^2\right)\;\;.
\eq
For this value of $\xi$ we then have
\bqa
X(\phi,\vec{z}) &=& F(\phi,\Xi(\phi,\vec{z}),\vec{z})\nl
&=& {3\over2}\phi - z_3\phi^2 - {5\over12}\phi^3
+ {1\over2z_q}\left(1-z_3\phi-{z_4\over2}\phi^2\right)^2\;\;.
\eqa
By the same arguments as before, we find that the leading asymptotic behaviour
is given by
\bqa
g(n_g,n_q;\vec{z}) & = & \suml_{m_{3,4,q}}
c(n_g,n_q;m_3,m_4,m_q)z_3^{m_3}z_4^{m_4}z_q^{m_q}\nl
&\sim& {1\over X(\hat\phi,\vec{z})^{n_g}\Xi(f,\vec{z})^{n_q}}\;\;,
\label{driez}
\eqa
where the saddle point $f$ is determined by
\bq
\left[{\partial\over\partial\phi}\left(\vphantom{A^A_A}
n_g\log X(\phi,\vec{z}) + n_q\log\Xi(\phi,\vec{z})\right)\right]_{\phi=f}\;\;.
\eq
From Eq.(\ref{driez}) it follows, by the probabilistic argument used before,
that for $n_g,n_q\to\infty$ the numbers $m_3$, $m_4$ and $m_q$ are
all normally distributed. The mean values and variances are again found by
taking the appropriate derivatives at $\vec{z}=(1,1,1)$. The saddle-point
value is, in that case, given by\footnote{There are other roots, but
since these do not depend on $\rho$ they cannot be relevant ones.}
\bq
R(\rho,f)\equiv 6(1-\rho)-12f + 3(1+\rho)f^2+(3-\rho)f^3 = 0\;\;,
\eq
where
\bq
n_q\equiv\rho N\;\;\;,\;\;\;N=n_g+2n_q\;\;.
\eq
The solution is given by
\bqa
f &=& {1\over3-\rho}\left\{
-1-\rho-2b
\sin\left[{1\over3}\arcsin\left(
{2\over b^3}\left(\rho^3-9\rho^2+15\rho-23\right)
\right)\right]\right\}\;\;,\nl
b &=& \left(13-2\rho+\rho^2\right)^{1/2}\;\;.
\eqa
The solution $f$ interpolates smoothly from $-1+\sqrt{3}\sim0.732$ at 
$\rho=0$ to $-3/5+14/5\sin(\arcsin(282/343)/3)\sim0.2854$
 at $\rho=1/2$. 
The results for the mean values are
\bqa
\langle m_q\rangle &=&
N{6+3f^2-12\rho f^2-12 f-18\rho f+3f^3+13\rho f^3+5\rho f^4\over
(1-f)(-24f+5f^3+9f^2+6)}\;\;,\nl
\langle m_3\rangle &=&
N{2f (18f-12\rho f-6f^3+7\rho f^3-9f^2+9\rho f^2-6+6\rho)\over
(-24f+5f^3+9f^2+6)(-2+f^2+2f)}\;\;,\nl
\langle m_4\rangle &=&
N{f^2(16f-8\rho f-5f^3+5\rho f^3-7f^2+5\rho f^2-6+6\rho)\over
(-24f+5f^3+9f^2+6)(-2+f^2+2f)}\;\;.
\label{QCDavgs}
\eqa
In figure 6 we give the results for the expectation
values $\langle m_j\rangle/N$, with $j=q,3,4$.
\begin{figure}[ht]
\begin{center}
\plaat{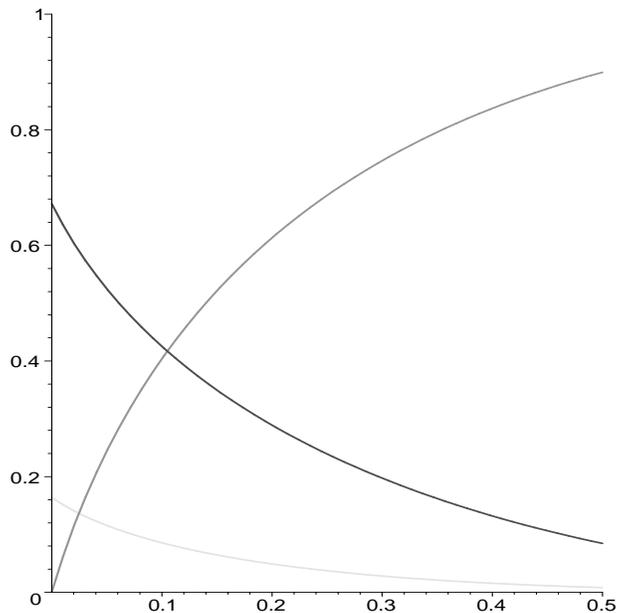}{8}{8}{0}
\caption[.]{Expected number of vertices of given type as a fraction of 
$N=n_g+2n_q$
in QCD with a single quark flavour. Along the horizontal axis we plot
$\rho=n_q/N$. 
The rising curve is the result for
the quark-gluon vertices, the falling curves are for the four-gluon
vertex (lowest curve) and the three-gluon vertex (middle curve).}
\end{center}
\end{figure}
As a final check, note that the topological sum rule for the vertices
in this case reads
\bq
m_3 + 2m_4 + m_q = n_g + 2n_q - 2 \sim N\;\;.
\eq
This is borne out by the numerical results, but also the analytic estimates
(\ref{QCDavgs}) show that the combination
$$
\langle m_3 + 2m_4 + m_q - N\rangle
$$
is precisely proportional to $R(f,\rho)$ and hence vanishes at the
saddle point.

\end{document}